# A Complexity measure based on Requirement Engineering Document

Ashish Sharma, D.S. Kushwaha

**Abstract** - Research shows, that the major issue in development of quality software is precise estimation. Further this estimation depends upon the degree of intricacy inherent in the software i.e. complexity. This paper attempts to empirically demonstrate the proposed complexity which is based on IEEE Requirement Engineering document. It is said that a high quality SRS is pre requisite for high quality software. Requirement Engineering document (SRS) is a specification for a particular software product, program or set of program that performs some certain functions for a specific environment. The various complexity measure given so far are based on Code and Cognitive metrics value of software, which are code based. So these metrics provide no leverage to the developer of the code. Considering the shortcoming of code based approaches, the proposed approach identifies complexity of software immediately after freezing the requirement in SDLC process. The proposed complexity measure compares well with established complexity measures. Finally the trend can be validated with the result of proposed measure. Ultimately, Requirement based complexity measure can be used to understand the complexity of proposed software much before the actual implementation of design thus saving on cost and manpower wastage.

**Index Terms**—Requirement Based Complexity, Input Output Complexity, Product Complexity, Personal Complexity Attributes, Interface Complexity, User Location Complexity, Requirement Complexity.

——————————— ◆ ———————————

## 1 INTRODUCTION

The software complexity can be defined as "the degree to which a system or component has a design or implementation that is difficult to understand and verify [5].

Most of the software complexity measures are based on code, but when we have the code for software, it is too late. In order to propose a comprehensive measure, we have taken the IEEE software requirement specification (SRS) document [1] as our foundation. Proper analysis of each and every component of SRS document has contributed towards finding out the complexity as shown in figure 1. Now these attributes in a procedural fashion have helped us to create a complexity measures based on requirement which in turn will be comparable to the code complexity metrics. The code and design decisions can be made much in advance. Further it will be cost effective and time saving.

Since all the established complexity measures for software like Halstead software difficulty metric [2], Mc Cabe Cyclometric complexity metrics [3], Klemola's KLCID complexity metric [6], Wang's cognitive functional complexity[11], Kushwaha's Cognitive Information Complexity Measure [5] and many more [7],[8] are code based but our proposed measure is based on Requirement Engineering document. Now, to generate a measure based on SRS following are the points which are considered during formation of this complexity measure

- Code based measures include details of the code and from that code we find out the measures like LOC, ID, OPERATORS, OPERAND, DIFFICULTY, V(G), CFS, CICM etc. but in these details are not available so proposed measure calculate the complexity by considering the requirements in detail so as to make the proposed complexity comparable with the code based complexities [2], [3]. Proposed measure also consider the same level of detailing as it is done in established measures.

- In the same line, we can decompose the requirements such that we can compare further with code. These requirements can be Functional and Non Functional. It may be appropriate to partition the functional requirement into sub function or sub processes [1]. This does not imply that the software design will also be portioned that way using this concept we can decompose our Functional Requirements

- In addition to this, attributes which are captured from SRS for finding out Req. based Complexity are given as

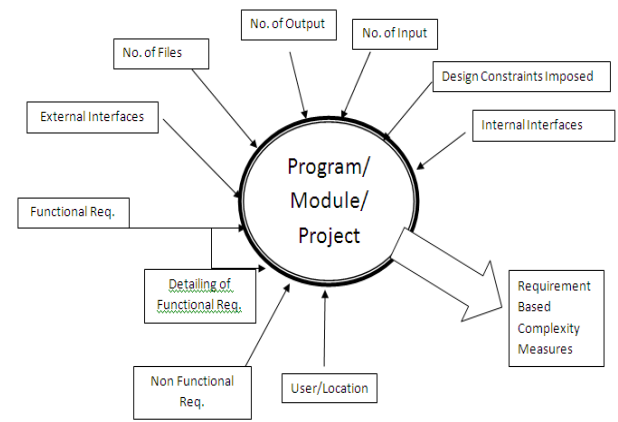

Figure 1. Factor derivation from SRS for RBC

————————————————

- *Ashish Sharma, Department of Computer Science & Engineering, Motilal National Institute of Technology, Allahabad.*
- *D.S. Kushwaha, Department of Computer Science & Engineering, Motilal National Institute of Technology, Allahabad.*

JOURNAL OF COMPUTER SCIENCE AND ENGINEERING, VOLUME 1, ISSUE 1, MAY 2010

## 2 ESTABLISHED COMPLEXITY MEASURES

As such, none of the complexity measure gives any method for computation of complexity based on requirements. Since this a new approach and to be compared with existing code based approaches for the validation of result. Following are the code based and Cognitive information based complexity measures.

### 2.1 CODE BASED COMPLEXITY MEASURES [2] [3]

#### 2.1.1 HALSTEAD COMPLEXITY MEASURE [2]

Maurice Halstead proposed this measure which is based on the principle of Count of Operators and Operand and their respective occurrences in the code. These operators and operands are to be considered for the formation of Length and Vocabulary of Program. Further Program Length and Vocabulary serve as basis for finding out Volume, Potential Volume, Estimated Program length, Difficulty and finally effort and time by using following formulae.

Program Vocabulary, $n = n1+n2$

Program Length, $N = N1+ N2$

Volume, $V= N*\log_2 n$

Estimated Program Length $N\hat{} = n1 \log_2 n1 + n2 \log_2 n2$

Potential Volume, $V* =(2+n2*)\log_2(2+n2*)$

Program Level, $L = V*/V$

Effort, $E =V/L$ in elementary mental discriminations

Reasonable Time, $T = E/B$ min

Difficulty = 1/language level

Now the problem with this method is that, they are difficult to compute. It is not suited when we want fast and easy computation, because to count distinct operand and operator is not easy job. Specifically when there are large programs.

#### 2.1.2 MAC CABE'S CYCLOMETRIC COMPLEXITY [3]

One of the better known and graphic metrics is Cyclometric Complexity developed by Thomas J Mc Cabb in 1976. His fundamental assumption was that software complexity is intimately related to the number of control paths generated by the code. The metric can be defined in two equivalent ways.

The number of decision statement in a program + 1

Or for a graph G with n vertices, e edges and p connected components,

$$v(G) = e-n+2p$$

Finally number of branches can be counted from the graph. The Mc Cabb complexity C can be defined as:

$$C = 1 + \sum_{n \subset G} (\text{degree } (n) - 1)$$

The difficulty with Mc Cabb Complexity is that, the complexity of an expression with in a conditional statement is never acknowledged. Also there is no penalty for embedded loops versus a series of single loops; both have the same complexity.

### 2.2 COGNITIVE COMPLEXITY MEASURES

#### 2.2.1 KLCID COMPLEXITY METRICS [6]

Klemola and Rilling proposed KLCID based complexity measure in 2004. It defines identifiers as programmer defined variables and based on identifier density (ID)

ID = Total no. of identifiers/ LOC

For calculating KLCID, it finds no. of unique lines of code, lines that have same type and kind of operands with same arrangements of operators would be consider equal. I defines KLCID as:

KLCID= No. of Identifier in the set of unique lines/ No. of unique lines containing identifier

This method can become very time consuming when comparing a line of code with each line of the program. It also assumes that internal control structures for the different software's are same.

#### 2.2.2 COGNITIVE FUNCTIONAL COMPLEXITY [5]

Wang and Shao have proposed functional size to measure the cognitive complexity. The measure defines the cognitive weights for the Basic Control Structures (BCS). Cognitive functional size of software is defined as:

$$CFS = (Ni + No) * Wc$$

Where Ni= No. of Inputs, No= No. of Outputs and Wc=Total Cognitive weight of software

Wc is defined as the sum of cognitive weights of its q linear block composed in individual BCS's. Since each block may consist of m layers of nesting and each layer with n linear BCS, total cognitive weight is defined as:

$$W_c = \sum_{i=1}^{q} \left\{ \prod_{k=1}^{m} \sum_{i=1}^{n} w_c\ (j, k, i) \right\}$$

Only one sequential structure is considered for a given component.

Now difficulty with this measure is that, it does not provide an insight into the amount of information contained in software.

#### 2.2.3 COGNITIVE INFORMATION COMPLEXITY MEASURE [4]

This measure is defined as product of weighted information count of the software and sum of the cognitive weights of Basic Control Structure (SBCS) of the software.

$$CICM = WICS * SBCS$$

This establishes a clear relationship between difficulty in understanding and its cognitive complexity. It also gives the measure of information contained in the software as:

$$Ei=ICS/ LOCS$$

where Ei represents Information Coding Efficiency.

The cognitive information complexity is higher for the programs, which have higher information coding efficiency.

Now the problem with these measures are that, they all uses code or in other words we can say that they are code dependent measures, which itself is a problem as stated earlier.

Various theories have been put forward in establishing code complexity in different dimensions and parameters.

## 3 REQUIREMENT BASED COMPLEXITY MEASURE

As described earlier in the paper that, this measure is based on the factors derived from SRS Document. The advantage of this approach is that it is able to estimate the software complexity in early phases of software life cycle, even before analysis and design is carried out. Due to this fact this is a cost effective and less time consuming. Now the calculation method for this measure based on different parameters is proposed next.

### COMPLEXITY ATTRIBUTE 1:

### INPUT OUTPUT COMPLEXITY (IOC)

This complexity refers to the input and output of the software system and attached interfaces and files. Following four attributes are considered:

Input: As Information entering to the System
Output: Information Leaving System
Interface: User Interface where the Input are to be issued and output to be seen and specifically number of integration required
Files: This refers to the data storage required during transformation

Now, Input Output Complexity can be defined as:

IOC = No. of Input + No. of Output + No. of Interfaces + No. of files      (1)

### COMPLEXITY ATTRIBUTE 2:

### FUNCTIONAL REQUIREMENT (FR)

Functional Requirement defines the fundamental actions that must take place in the software in accepting and processing the inputs and in processing and generating outputs. Functionality refers to what system is supposed to do. This describes the general factor that affects the product and its functionality. Every stated requirement should be externally perceivable by users, operators or other external systems.

It may be appropriate to partition the functional requirement into sub-functions or sub-processes

FR = No. of Functions * $\sum_{i=1}^{n} SPFi$      (2)

Where SPF is Sub Process or Sub-functions received after decomposition.

### COMPLEXITY ATTRIBUTE 3:

### NON FUNCTIONAL REQUIREMENT (NFR)

This refers to the Quality related requirements for the software apart from functionality. These requirements are categorized into THREE categories with their associated precedence values as shown in Table 1. As high the precedence that much high will be the value, which will further depend upon the count. It can be mathematically described as:

NFR = $\sum_{i=1}^{3} \sum_{j=1}^{3} (Type_i * Count_j)$      (3)

Table 1- Describes the different types of Non Functional Req

| Type | Count |
|---|---|
| Optional Req. | 1 |
| Must be Type | 2 |
| Very Important Type | 3 |

### COMPLEXITY ATTRIBUTE 4:

### REQUIREMENT COMPLEXITY (RC)

It refers to the sum of all requirements i.e. ffunctional and its decomposition into sub-functions and non ffunctional requirements:

RC = FR * NFR      (4)

### COMPLEXITY ATTRIBUTE 5:

### PRODUCT COMPLEXITY

This refers to the overall complexity based on its functionality of the system. We have proposed this a product of Requirement Complexity and Input Output Complexity. It can be mathematically described as:

PC = IOC * RC      (5)

### COMPLEXITY ATTRIBUTE 6:

### PERSONAL COMPLEXITY ATTRIBUTES:[9] [10]

For effective development of software, Technical Expertise plays a very significant role. Now computation of the Personal Attributes lead to technical expertise, and this is referred to as the "Multiplier Values for Effort Calculation i.e. Cost Driver Attributes of Personal Category from COCOMO Intermediate model proposed by Berry Boehm and they are shown as follows

Table 2: Cost Driver Attributes and their values used in COCOMO Model

| Attribute | Rating | | | | |
|---|---|---|---|---|---|
| | Very Low | Low | Nominal | High | Very High |
| Analyst Capability | 1.46 | 1.19 | 1.00 | 0.86 | 0.71 |
| Application Exp. | 1.29 | 1.13 | 1.00 | 0.91 | 0.82 |
| Programmer Cap. | 1.42 | 1.17 | 1.00 | 0.90 | -- |
| Virtual Machine Exp. | 1.21 | 1.10 | 1.00 | 0.90 | -- |
| Programming Language Exp. | 1.14 | 1.07 | 1.00 | 0.95 | -- |

Mathematically PCA can be described as Sum of Product of attributes as mentioned in above table.

PCA = $\sum_{i=1}^{5} Mf$      (6)

where Mf are Multiplying Factor.

## COMPLEXITY ATTRIBUTE 7:

### DESIGN CONSTRAINTS IMPOSED (DCI)

It refers to no. of cconstraints that are to be considered during development of software/ system by any statuary body/ agencies which includes number of regulatory constraints, hardware constraints, communication constraints, database constraints etc. This metrics can be mathematically defined as.

$$DCI = \sum_{i=0}^{n} C_i \qquad (7)$$

Where Ci is Number of Constraints and value of Ci will vary from 0 to n.

$$C_i = \begin{cases} 0 & \text{If Blind Development} \\ \text{Value} & \text{If Constraints exists} \end{cases}$$

## COMPLEXITY ATTRIBUTE 8:

### INTERFACE COMPLEXITY

This complexity attribute is used to define number of External Integration/Interfaces to the proposed module/ program/ system. These interfaces can be hardware interface, communication interface and software interface etc.

$$IFC = \sum_{i=0}^{n} EI_i \qquad (8)$$

Where EIi is Number of External Interfaces and value of EIi will vary from 0 to n

$$IFC = \begin{cases} 0 & \text{If no External Interface} \\ \text{Value} & \text{If External Interface exists} \end{cases}$$

## COMPLEXITY ATTRIBUTE 9:

### USERS/ LOCATION COMPLEXITY

This measure refers to the number of user for accessing the system and locations (Single or Multiple) on which the system is to be deployed/ used

$$ULC = \text{No. of User} * \text{No. of Location} \qquad (9)$$

## COMPLEXITY ATTRIBUTE 10:

### SYSTEM FEATURE COMPLEXITY

This refers to the specific features to be added to the system so as to enhance look and feel feature of the system

$$SFC = (Feature1 * Feature2 * \ldots\ldots\ldots * Feature\ n) \qquad (10)$$

## COMPLEXITY DEFINITION:

### REQUIREMENT BASED COMPLEXITY

Finally the Requirement Based Complexity can be obtained by considering all above definitions. It can be mathematically shown as:

$$RBC = ((PC * PCA) + DCI + IFC + SFC) * ULC \qquad (11)$$

The Requirement Based Complexity will be higher for the programs, which have higher Functionality to be performed and more quality attributes which is to be retained. All above measure have been illustrated with the help of an example below:

Example # 1

Consider a program to be developed for finding out the factorial of a given number. Upon going through the SRS, we are able to extract the following parameters.:

| | |
|---|---|
| Number of Inputs | 01 (Number) |
| Number of Outputs | 01 (Factorial of a number) |
| Number of Interfaces | 01 (User Interface) |
| Number of Files | 01 (For storage of Values) |

IOC = 1+1+1+1 = 4

| | |
|---|---|
| Number of Functional Req. | 01 (Factorial) |
| Number of Sub-processes | 02 (Multiply, Decrement) |

FR = 1 * 2 = 2

No. of Non FR    00 (no Quality attribute)

RC = FR + NFR = 02

Product Complexity, PC = (IOC * RC) = 8

PCA = 1.17 (Suppose Programmer Capability = Low)
No. of Constraints    00 (No directives)
DCI= 0
IFC = 0;
Since this program is not to be further connected with any external interface
No. Of User and Location, ULC = 1 * 1= 01
Now, RBC = ((PC * PCA) + DCI + IFC + SFC) * ULC
**Requirement Based Complexity = 9.36**

The complexity measured by us for the given SRS, the program code is illustrated in figure 2. Based on the above code we compute the complexity of the other proposed measures.

Figure 2: Program code for Factorial of a number

```
#include<stdio.h>
#include<conio.h>
int factorial(int);
void main()
{
 int n, fact;
 clrscr();
 printf("\n\t\t Program to calculate the factorial of a number:");
 printf("\n\n\t\t Enter number to calculate the factorial:");
 scanf("%d",&n);
 fact =factorial(n);
 printf("Factorial of %d = %d",n,fact);
 getch();
}
 int factorial( int n )
 {
  if ( n == 0 )
   return 1;
  else
   return n* factorial(n-1);
 }
```

Now calculation of other measures as:

| KLCID | | CFS | | CICM | | Mc Cabb | | Halstead | |
|---|---|---|---|---|---|---|---|---|---|
| ID | 9 | (Ni) | 1 | LOC | 19 | Nodes | 19 | n1 | 18 |
| LOC | 19 | (No) | 1 | Identifier | 9 | | | n2 | 4 |
| ID | 2.11 | BCS (Seq) | 1 | | | Edges | 20 | N1 | 35 |
| No. of Unique Lines containing identifier | 5 | BCS (func call) | 2 | SBCS | 3 | Predicate Node | 2 | N2 | 10 |
| | | | | | | | | Vocabulary | 22 |
| No. of identifier in the set of Unique Lines | 7 | Wc | 3 | WICS | 5.33 | Regions | 3 | Pgm. Length | 45 |
| | | | | | | | | Effort | 5154 |
| | | | | | | | | Time | 4.78 |
| KLCID | 1.4 | CFS | 6 | CICM | 16.01 | V(G) | 3 | Halsted Difficulty | 22.5 |

## 4 RESULT

This section analyses the result of applying RBC on 16 programs selected and developed on a mix of language like C, C++, Java. So as to find the complexity variation in terms of code. In order to analyze the validity of the result, the RBC for different program is calculated based on SRS and further compared with other established measures which are based on Code and Cognitive complexity.

Further based on these values a plot for Requirement based complexity versus Other Established Measures (Code Based & Cognitive Complexity Based) is plotted and observed that all the values are aligned.

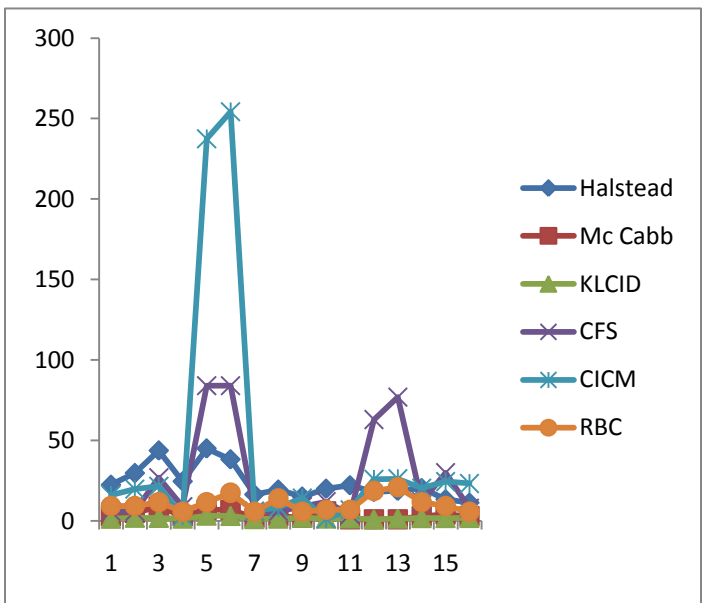

*Figure 4*

## 5.0 CONCLUSION

This paper has developed the Requirement based complexity measure that is based on SRS document. It is a robust method because it encompasses all major parameters and attributes that are required to find out complexity. Further they are comparable with code based and cognitive information based complexity measures. On comparing the Requirement based complexity measure with rest of the established measure following are the findings:

i. RBC follows Code based measures which have been computed on the basis of Program by identifying number of operators and operands and further Vocabulary, Length and finally it is aligned with the Difficulty metrics given by Maurice Halsted.

ii. RBC also aligned with the Control Flow based/ Graphical complexity measure proposed by Thomas J Mc Cabb which was identified by creating Control Flow Graph for the given programs.

iii. Finally RBC is also giving the similar kind of result trend what the Cognitive Based Complexity measures are giving, which, can be computed considering software as information and drawing conclusion based on cognitive science theory where Identifier Density, Basic Control Structure, Cognitive weights and other measures have been taken for further exercise.

Finally we can say that proposed measure follow the trend of all the other established measure in comprehensive fashion. This measure is computationally simple and will aid the developer and practitioner in evaluating the software complexity in early phases which otherwise is very tedious to carry out as an integral part of the software planning. Since entire approach is based on Requirement Engineering document so it is for sure that an SRS must have all the characteristics, content and functionality to make this estimation precise and perfect. The method explained above is well structures and forms a basis for estimation of software complexity for early design decisions and implementation which is to be carried forward.

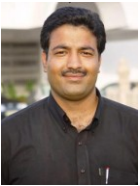

**Ashish Sharma** received his M.Tech. Degree in Comp. Sc. & Engg from U.P. Technical University, Lucknow, India in the year 2006. Currently he is pursuing Ph.D. in Comp. Sc. & Engg from MNNIT, Allahabad, India under the guidance of Dr. D.S. Kushwaha. He is presently working with the GLA Institute of Technology & Management, Mathura as Reader in the department of Computer Science & Engineering having experience of 12 years. His research interests include areas in Software Complexity, Software Testing and TQM. He is presently working on Requirement Based Complexity theory and Test Effort Estimation.

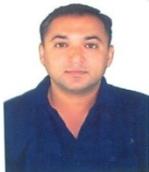

**Dr. D.S.Kushwaha** received his Doctorate Degree in Computer Science & Engineering from Motilal Nehru National Institute of Technology, Allahabad, India in the year 2007 under the guidance of Dr. A.K. Misra. He is presently working with the same Institute as Assistant Professor in the department of Computer Science & Engineering. His research interests include areas in Distributed Systems, High Performance computing, Service Oriented Architecture and Data Structures. He has over 35 International publications in various Conferences & Journals.